\documentclass[aps,prb,twocolumn,groupedaddress,floats,showpacs]{revtex4}
\usepackage{latexsym}
\usepackage{dcolumn}
\usepackage[dvips]{graphicx}
\usepackage{amssymb}
\usepackage{graphics}
\usepackage{amsmath}
\usepackage{epsf}

\begin{document}

\author{S. Adam, E. H. Hwang, E. Rossi, and S. Das Sarma }
\affiliation{Condensed Matter Theory Center, Department of Physics,
University of Maryland, College Park, MD 20742-4111}
\title{Theory of charged impurity scattering in two dimensional graphene}
        
\date{\today}
\begin{abstract}
We review the physics of charged impurities in the vicinity of
graphene.  The long-range nature of Coulomb impurities affects both
the nature of the ground state density profile as well as graphene's
transport properties.  We discuss the screening of a single Coulomb
impurity and the ensemble averaged density profile of graphene in the
presence of many randomly distributed impurities.  Finally, we discuss
graphene's transport properties due to scattering off charged
impurities both at low and high carrier density.
\end{abstract}
\maketitle
 
 \section{Introduction}
 
 Graphene is a two dimensional sheet of carbon whose atoms arrange in a
 honeycomb lattice with nearest neighbor atoms forming strong sp$_2$
 bonds.  The electronic properties of this material are mostly
 determined by the $p_z$ orbitals with each carbon atom contributing
 one electron to a Bloch band whose low energy properties are
 adequately described by a Dirac-Weyl effective Hamiltonian.  While
 the study of Dirac Fermions has emerged in several contexts in
 theoretical condensed matter physics, its experimental realization
 about three years ago, in the form of gated graphene
 devices,~\cite{kn:novoselov2004,kn:novoselov2005,kn:zhang2005} where
 the carrier density can be tuned continuously from electron-like
 carriers for positive bias to hole-like carriers for negative gate
 voltage, has prompted a prolific theoretical and experimental effort
 to understand the properties of this novel material.

Most of excitement surrounding graphene stems from one of the
following peculiar properties: (i) Electrons and holes in graphene
have a gapless linear dispersion relation in contrast to the parabolic
dispersion of other more conventional electron gases; (ii) The
carriers in graphene are chiral -- a property that has
striking consequences such as the ``half-integer'' quantum Hall
Effect~\cite{kn:novoselov2005,kn:zhang2005}; and (iii) Carriers in
graphene live at an exposed almost perfect $2$D surface that is
amenable to surface
probes~\cite{kn:ishigami2007,kn:stolyarova2007,kn:rutter2007,kn:li2007,kn:martin2008}
and surface manipulation.~\cite{kn:schedin2006,kn:jang2008}  In
addition, we note that there is the potential of mass producing
graphene through epitaxial growth methods,~\cite{kn:berger2006} and
that graphene has remarkable mechanical
properties~\cite{kn:bunch2007,kn:lee2008} which only further enhance
the interest.

\begin{figure}[t!]
\bigskip
\hspace{0.1\hsize}
\epsfxsize=0.9\hsize
\epsffile{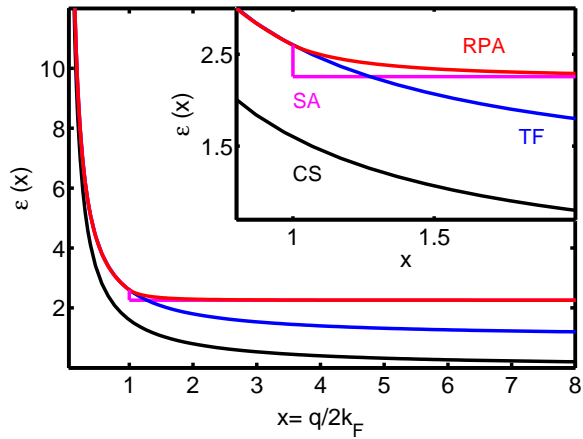}
\caption{\label{Fig:dielectric_function} 
The main panel shows different dielectric functions used in the 
literature, including the ``Complete Screening" (CS),``Thomas-Fermi" (TF)
and ``Random Phase Approximation" (RPA).  The inset shows
a blow-up at $q=2 k_{\rm F}$ to show where ``Step Approximation" (SA) 
used in Ref.~\onlinecite{kn:adam2007a} differs from the exact result. }
\end{figure}

In this Perspective, we look at one important aspect of graphene which
 is the influence of disorder on its ground state and transport
 properties.  We demonstrate that for graphene, charged (i.e. Coulomb) impurities
 behave qualitatively different from neutral
 impurities~\cite{kn:aleiner2006,kn:altland2006,kn:foster2008} and dominate
 graphene's transport properties at low carrier density.  The importance of
 the Coulomb nature of graphene impurities was first highlighted by
 Ando,~\cite{kn:ando2006} where by calculating the intraband
 contribution to the polarizability and absorbing
 the interband (i.e. electron-hole)
contribution  
into a redefinition of the dielectric
 constant~\cite{kn:gonzalez1994} he showed that charged impurities could
 explain the conductivity being linear-in-density as was seen in
 experiments.~\cite{kn:novoselov2004,kn:tan2007}  Similar conclusions
 were obtained by Nomura and
 MacDonald~\cite{kn:nomura2006a} using a ``complete
 screening" model (i.e. $r_s \rightarrow \infty$), Cheianov and Falko
 using a numerical Thomas-Fermi approximation~\cite{kn:cheianov2006}
 and in Ref.~\onlinecite{kn:hwang2006c} using the full
 Random-Phase-Approximation (RPA).  Analytic expressions for the RPA
 polarizability function calculated first in
 Ref.~\onlinecite{kn:hwang2006b} and then in
 Refs.~\onlinecite{kn:wunsch2006,kn:barlas2007,kn:polini2007} revealed
 that for momentum transferred on the Fermi circle (i.e. $q = |{\mathbf
 k} - {\mathbf k}' |= 2 k_{\mathrm F} \sin\theta/2 \leq 2 k_{\mathrm F}$) the graphene dielectric
 function calculated using the RPA was identical to the much simpler
 Thomas-Fermi approximation at $T=0$ (see Fig.~\ref{Fig:dielectric_function}).  This then made it possible
 to calculate the RPA-Boltzmann conductivity
 analytically,~\cite{kn:adam2007a} and the dependence of graphene's 
 conductivity on the fine-structure constant $r_s 
\equiv e^2/(\hbar v_{\rm F} \kappa)$ was recently 
 verified experimentally.~\cite{kn:jang2008} The importance of Coulomb 
 scattering in explaining the observed
 graphene transport properties soon prompted an interest in
 investigating the properties of a single charged impurity embedded in
 graphene.  Katsnelson~\cite{kn:katsnelson2006} studied this problem
 using a Fermi-Thomas approximation, followed by studies in
 Refs.~\onlinecite{kn:shytov2007,kn:pereira2007,kn:biswas2007,kn:novikov2007,kn:fogler2007,kn:terekhov2008,kn:pereira2008}
 who were mostly interested in effects beyond the RPA such as
 determining the critical impurity charge for which the Coulomb
 impurity forms bound states and the screening properties of graphene
 in the supercritical regime.  

It was understood by Refs.~\onlinecite{kn:hwang2006c,kn:nomura2007}
that as one approached the Dirac point, one would soon encounter a
situation where the gate voltage induced carrier density would be
smaller than the fluctuation of carrier density induced by the charged 
impurities thereby breaking the graphene landscape into puddles of electrons and holes.
Solving numerically for the conductivity using a
finite-sized Kubo formalism for a limited range of impurity concentrations, 
Ref.~\onlinecite{kn:nomura2007} concluded that the Coulomb disorder model gave a
{\it universal} minimum conductivity whose value did not depend on the
charged impurity concentration, but that was larger than that expected
for clean Dirac Fermions,~\cite{kn:fradkin1986,kn:ludwig1994,kn:shon1998,kn:katsnelson2006,kn:tworzydlo2006}
while Ref.~\onlinecite{kn:hwang2006c} argued that this would give rise to a  {\it
non-universal} minimum conductivity whose value depended on the
concentration of charged impurities.  Ref.~\onlinecite{kn:adam2007a} developed a mean 
field approach to
understand the properties of graphene at the Dirac point by
calculating an effective carrier density self-consistently.  This
theory made quantitative predictions about the dependence of the
minimum conductivity and {\rm rms} carrier density on the charged
impurity concentration and substrate dielectric constant, and in
particular argued that cleaner graphene samples would have larger
minimum conductivity.  Ref.~\onlinecite{kn:rossi2008} then studied the
ground state properties of graphene by minimizing an energy functional
comprising kinetic energy, Hartree, exchange~\cite{kn:hwang2007,
kn:barlas2007, kn:dassarma2007b} and correlation~\cite{kn:barlas2007,kn:dassarma2007b,kn:polini2008}
contributions in the presence of Coulomb disorder.  This work made
quantitative predictions about properties of the carrier density
distribution, both at and away from the Dirac point, and enabled
Ref.~\onlinecite{kn:rossi2008b} to develop an effective medium theory
to calculate the graphene's conductivity through these inhomogeneous
puddles, capturing quantitatively the minimum conductivity plateau
that is seen in
experiments.~\cite{kn:novoselov2004,kn:tan2007,kn:chen2008}  We
mention that underlying the existence of this minimum conductivity
plateau is the high transmission of graphene p-n junctions, which has
been the subject of theoretical~\cite{kn:cheianov2006b,kn:fogler2008}
and experimental
study.~\cite{kn:huard2007,kn:williams2007,kn:oezyilmaz2007} For
the purposes of this paper we do not discuss quantum interference
effects (see Ref.~\onlinecite{kn:adam2008d} and references therein) or
the strongly interacting regime (see Ref.~\onlinecite{kn:muller2008}
and references therein).

The remainder of this paper is structured as follows.  In
Section~\ref{Sec:Single} we discuss the problem of the screening of a
single Coulomb impurity in the sub-critical regime as a useful toy
model to understand the many impurity problem that we address in
Section~\ref{Sec:Ground} where we study the case of many Coulomb
impurities that are uncorrelated and distributed uniformly in order to
study the ground state properties of graphene.  In
Section~\ref{Sec:Boltzmann}, we review the high-density Boltzmann
transport theory, and discuss the Effective Medium Theory (EMT) in
Section~\ref{Sec:EMT}. In 
Sections~\ref{Sec:Expts},~\ref{Sec:MinCond}, and~\ref{Sec:Recent} 
we briefly review the experimental situation,  discuss 
graphene minimum conductivity, and recent 
theoretical work not covered in this review. We then conclude in
Section~\ref{Sec:Conclusion}.   

 \section{Screening of a single Coulomb impurity}
 \label{Sec:Single}
 
Following Ref.~\onlinecite{kn:katsnelson2006} one can construct the
Thomas-Fermi screening of a single charged impurity.  The goal is to
calculate the screened Coulomb potential $V_s(r) = V_0(r) + V_{\rm
ind}(r)$, where the bare potential $V_0 = \hbar v_{\rm F} r_s [r^2 +
d^2]^{-1/2}$ and the induced potential is given by
\begin{equation}
V_{\rm ind}(r) = (\hbar v_{\rm F} r_s) \int d{\bf r}'
\left. \frac{n(r') - {\bar n}}{|{\bf r} - {\bf r}'|} \right.,
\end{equation}
where we can imagine tuning the back gate to ensure charge neutrality
${\bar n} = 0$.  If one further assumes that the local carrier density
is given by the Fermi-Thomas condition, $n[V(r)] = -V(r)^2/[\pi (\hbar
v_{\rm F})^2]$, one can write down a (one dimensional)
self-consistency equation for ${\tilde V_s} = V_s/\hbar v_{\rm F} r_s$ 
\begin{equation}
{\tilde V}_s(r) = \frac{1}{\sqrt{r^2 + d^2}} 
- \frac{4 r_s^2}{\pi} \int dr'  \left. \frac{r'}{r+r'}
K\left[\frac{4 r r'}{(r+r')^2}\right] {\tilde V}_s^2(r') \right.,
\end{equation}
where $K[x]$ is the complete elliptic integral of the first kind.  
The screened
potential induced for this single impurity using this method
was discussed in Ref.~\onlinecite{kn:katsnelson2006}.  This formalism
can be generalized using the method developed 
in Ref.~\onlinecite{kn:rossi2008} to include the effects of exchange.  
The ground state carrier density can be obtained from the 
Thomas-Fermi-Dirac (TFD) energy functional
\begin{eqnarray}
 E[n] &=& \hbar v_F \left[ \frac{2\sqrt{\pi}}{3}\int d^2 r 
         \mbox{sgn}(n)|n|^{3/2} \right.  \nonumber \\
	&& \mbox{}  + \left. \frac{r_s}{2} \int  d^2 r
          \int d^2 r' \frac{n({\bf r})n({\bf r}' )}{|{\bf r} - {\bf r}'|} 
         +\frac{E_{xc}[n]}{\hbar v_F}  \right.\nonumber \\ 
          && \mbox{} \left.   +r_s\int d^2 r V_D({\bf r})n({\bf r}) - 
          \frac{\lambda}{\hbar v_F}\int d^2 r n({\bf r})\right]
 \label{Eq:en}
\end{eqnarray}
where the first term in Eq.~\ref{Eq:en} is the kinetic energy, the
second term is the Hartree part of the Coulomb interaction, the third
is the exchange-correlation energy and the fourth term is the energy
due to disorder, where $V_D$ is the disorder potential and the last
term is added to set the average carrier density, $\langle n\rangle$,
through the chemical potential $\lambda$.  The correlation term is much
smaller than exchange and, to very good approximation
\cite{kn:barlas2007,kn:dassarma2007b,kn:polini2008} is proportional to exchange.
Therefore, hereafter, we neglect the correlation contribution by
assuming ${\delta E_{xc}}/{\delta n} = \Sigma(n)$, where $\Sigma(n)$
is the Hartree-Fock self-energy.~\cite{kn:barlas2007,kn:dassarma2007b,kn:hwang2007}
The energy functional Eq.~\ref{Eq:en} is quite general and can be
tailored by properly choosing $V_D$ and its coupling to $n({\bf r})$,
to consider different sources of disorder.  For
the single impurity problem, the solution can also be cast as a
 one dimensional integral equation 
\begin{eqnarray}
\frac{V_s(r)}{\hbar v_{\rm F}}
&=& \frac{r_s}{\sqrt{r^2 + d^2}}  \nonumber \\
&& \mbox{} + 
4 r_s  \int dr' \left. \frac{r'}{r+r'}
K\left[\frac{4 r r'}{(r+r')^2}\right] n(r') \right. \nonumber \\
&=& - {\rm sgn}(n) \sqrt{\pi |n(r)|} 
\left[
1 + \frac{1}{4} \ln\left(\frac{4 \Lambda}{\sqrt{\pi |n(r)|}} \right)
 \right. \nonumber \\
&& \mbox{} + \left.
r_s \left(\frac{2C + 1}{2 \pi} + \frac{1}{8}\right) \right],
\end{eqnarray}   
where $\Lambda = 1/(0.25~{\rm nm})$ is the band energy cutoff
and $C \approx 0.916$.  

% \begin{figure}
% \bigskip
% \hspace{0.1\hsize}
% \epsfxsize=0.9\hsize
% \epsffile{temp2.eps}
% %\includegraphics[scale=0.3]{fig_ssc2.eps}
% \caption{\label{Fig:single} Single impurity problem
% in graphene.  Plot shows comparison between 
% non-linear Thomas-Fermi and the solution
% from minimizing graphene's energy functional.}
% \end{figure}

 \section{Ground state properties at the Dirac point}
 \label{Sec:Ground}

The single impurity problem discussed in the previous section is a 
much simpler
problem because rotational symmetry makes the problem 
one-dimensional.  Adding many impurities also brings
further complications: while the carrier density induced by a single
impurity is negligible, this is not the case for many impurities where
although the average density can be tuned to zero via an external gate
potential, the scale of the density fluctuations is set 
by the impurity concentration.~\cite{kn:hwang2006c}  As shown
in Fig.~\ref{Fig:dielectric_function}, the RPA screening properties of
graphene are very different at the Dirac point (i.e. $k_{\rm F} \rightarrow 0$)
and at finite density, therefore theoretical frameworks constructed 
to work at the Dirac point are bound to fail when there are such 
large density fluctuations.  In this section, we present two different
approaches to describe the Dirac point.  The first is a mean-field 
theory where an effective density $n^*$ is obtained by solving
self-consistently for the density induced by the fluctuations of the 
screened impurity potential (that itself depends on density).  The second
is a generalization of the energy functional method discussed above
for a single impurity to the much more complicated case of many Coulomb
impurities. 
        
 \subsection{Self-consistent Approximation (SCA)}
 \label{SubSec:SCA}
 
\begin{figure}
\bigskip
\hspace{0.1\hsize}
\epsfxsize=0.9\hsize
\epsffile{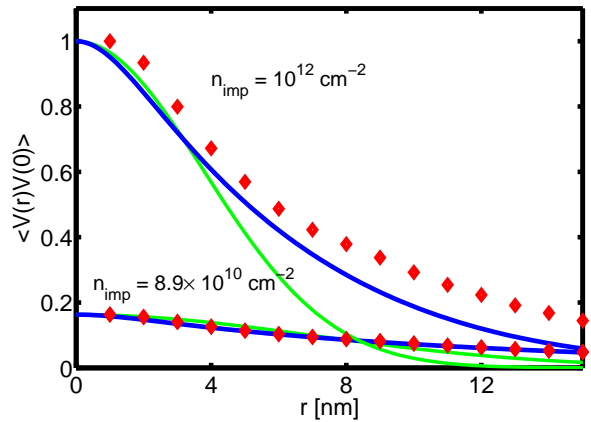}
  \caption{
           Spatial correlation function for the screened potential
	   at the Dirac point for $r_s=0.8$ and $d=1$~nm.
	   Red diamonds are the results obtained by minimizing graphene's
           energy functional
	   (Sec.~\ref{SubSec:EFM}).  The lines are the SCA results using 
	   Eq.~\ref{Eq:Corr1}, blue lines, and its Gaussian approximation,
	   Eq.~\ref{Eq:Corr2}. 	   
	   All the results are normalized via the value of $\langle V(0)V(0)\rangle$
	   for $n_{imp}=10^{12}$~cm$^{-2}$.
          }
  \label{fig:vsc}
\end{figure}

 For any microscopic single impurity potential $\phi(r,n)$, the
 probability distribution, $P(V )$, of the total potential, $V$ , is
 $P(V ) = \langle \delta (V - \sum_{i = 1}^{N_{\rm imp}} \phi(r_i,n))
 \rangle_{r_i}$ where $\langle \cdots \rangle_{r_i}$ is the average
 over all possible disorder configurations.  Assuming that the
 impurities positions are uncorrelated one can compute expressions for
 all moments of the induced disorder
 potential.~\cite{kn:galitski2007,kn:adam2008a} For example, the
 connected moment $\langle V^k \rangle_c = n_{\rm imp} \int d^2 r
 [\phi[(r,n)]^k$.  The self-consistent approximation involves
 obtaining the effective carrier density $n^*$ by equating the second
 moment of the disorder potential with the square of the corresponding
 Fermi energy $\langle V^2 \rangle_c = (E_{\rm F}[n^*])^2 = \pi (\hbar
 v_{\rm F})^2 n^*$.  This self-consistent approximation then allows us
 to compute any correlation function at the Dirac point, although
 closed form analytic results are often elusive.  To make analytical
 progress, one can map

\begin{subequations}
\begin{eqnarray}
\langle V(r) V(0) \rangle &=& n_{\rm imp} \int d {\bf q} 
[\phi(q, n^*)]^2 e^{i {\bf q} \cdot {\bf r}} \label{Eq:Corr1} \\
&&  \hspace{-1.1in} \approx \frac{n_{\rm imp} (\hbar v_{\rm F})^2 K_0[r_s, d \sqrt{n^*}]}
{2 \pi (\xi[r_s, d \sqrt{n^*})^2 } \exp \left[\frac{- n_{\rm imp} r^2}{2 (\xi[r_s, d \sqrt{n^*}])^2} \right],
\label{Eq:Corr2}
\end{eqnarray}    
\end{subequations}
where analytic expressions for $K_0$ and $\xi$, were reported in Ref.~\onlinecite{kn:adam2008c}.
A numerical evaluation of Eq.~\ref{Eq:Corr1} and the Gaussian approximation Eq.~\ref{Eq:Corr2} 
is shown in Fig.~\ref{fig:vsc}.  Within the Gaussian approximation one finds that 
$n_{\rm rms} = \sqrt{\langle V^4 \rangle}/[\pi(\hbar v_{\rm F})^2] \approx n^* \sqrt{3 + [n_{\rm imp} \pi \xi^2]^{-1}} 
\approx \sqrt{3} n^*$, where in the last equation we further assume that 
$n_{\rm imp} \pi \xi^2 \sim r_s^{-4} \gg 1$.  This result for $n_{\rm rms}$ is particularly useful when comparing the self-consistent approximation with other methods.

 \subsection{Energy Functional Minimization (EFM)}
 \label{SubSec:EFM} 

\begin{figure}
\bigskip
\hspace{0.1\hsize}
\epsfxsize=0.9\hsize
\epsffile{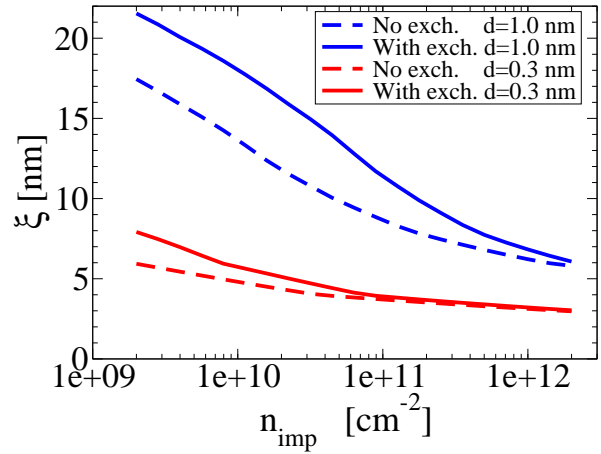}
 \caption{ \label{fig:xi} Density correlation length $\xi$ at the
 Dirac point as function of $n_{\rm imp}$ for $r_s=0.8$ and two
 different values of $d$.  Results with (without) exchange are shown
 by solid (dashed) lines.}
\end{figure}

\begin{figure}
\bigskip
\hspace{0.1\hsize}
\epsfxsize=0.9\hsize
\epsffile{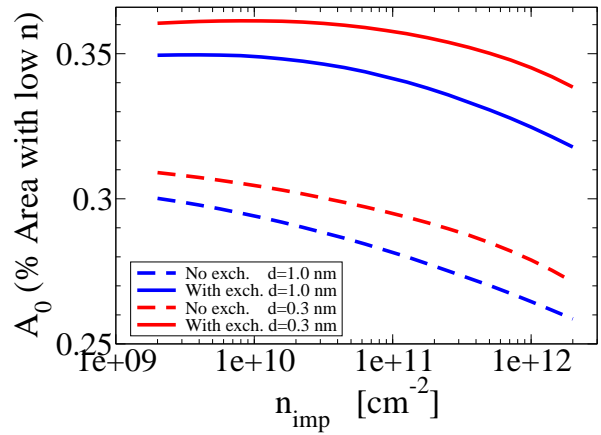}
 \caption{\label{fig:A0} Area fraction, $A_0$, over which is $|n({\bf
	   r})-\langle n\rangle|<n_{rms}/10$ as a function of
	   $n_{imp}$ at the Dirac point for $r_s=0.8$.  Results with
	   (without) exchange are shown by solid (dashed) lines.}
\end{figure}

 To study graphene transport properties for a distribution of charged
 impurities, we use Eq.~\ref{Eq:en} taking $V_D$ to be the 
potential generated by a 
 random 2D distribution $C({\bf r})$
 of impurity charges placed at a distance $d$ from the graphene
 layer. We assume $C({\bf r})$ to be on average zero and uncorrelated
 and perform our calculations on a 200~nm$\times$200~nm square sample
 with a 1~nm spatial discretization.  Close to the Dirac point, for a
 single disorder realization, we find that the carrier density breaks
 up into electron-hole puddles.  Since we are interested in disorder
 averaged quantities, we examine several disorder realizations
 (500-1000) and denote disordered averaged quantities by angled
 brackets.  To characterize the density profile, we
 calculate the disorder averaged density-density 
correlation function $\langle\delta
 n({\bf r})\delta n(0)\rangle$, from which we can extract the root
 mean square $n_{rms}=\sqrt{\langle\delta n(0)\delta n(0)\rangle}$,
 and the typical correlation length, $\xi$, defined in 
 this section as the FWHM of
 $\langle\delta n({\bf r})\delta n(0)\rangle$.  We find~\cite{kn:rossi2008} 
for typical graphene samples, that $n_{rms}\approx \langle n\rangle$ for
 dopings as high as $10^{12}$~cm$^{-2}$ and that close to the Dirac
 point, for $n_{imp}\lesssim 10^{10}$~cm$^{-2}$, $n_{rms}$ including
 exchange is three times smaller than without.  We find the
 correlation length $\xi$ to be of the order of $10$~nm, see
 Fig.~\ref{fig:xi}. This value suggest that the electron-hole puddles
 are quite small. However a closer inspection reveals that, close to
 the Dirac point, the density profile is characterized by two distinct
 types of inhomogeneities~\cite{kn:rossi2008b}: wide regions (i.e. big
 puddles spanning the system size) of low density containing a number
 of electrons (holes) of order 10; and few narrow regions, whose size
 is correctly estimated by $\xi$, of high density containing a number
 of carriers of order 2.  This picture is confirmed by the results
 shown in Fig.~\ref{fig:A0} in which the disorder averaged area
 fraction, $A_0$, over which $|n({\bf r})-\langle
 n\rangle|<n_{rms}/10$ is plotted as a function of $n_{imp}$. We see
 that $A_0$ is of order 1/3 and we also find that the area fraction
 over which $|n({\bf r})-\langle n\rangle|$ is less than $1/5$ of
 $n_{rms}$ is close to 50\% for $n_{imp}\lesssim 10^{11}$~cm$^{-2}$.
 The combination of the relatively high density in the peaks/dips and
 the fact that in the low density regions $n({\bf r})$ varies over
 scales much bigger than $10~{\rm nm}$ guarantees that the inequality
 $\sqrt{\pi n}[|\nabla n|/n]^{-1}\gg 1$ is satisfied over the majority
 of the graphene sample and therefore justifies the use of the EFM
 theory.  The EFM should be a reasonable quantitative theory for
 existing graphene samples at all values of the carrier density.

\subsection{Comparison of SCA and EFM}

Here we compare the results from the Self-Consistent Approximation of
Sec.~\ref{SubSec:SCA} and the Energy Functional Minimization (EFM)
formalism discussed in Sec~\ref{SubSec:EFM}.  Figure~\ref{fig:vsc}
shows the disordered averaged spatial correlation function at the
Dirac point for the screened disorder potential $V= V_D + (1/2)\int
d^2 r' n({\bf r}')/|{\bf r}-{\bf r}'|$ where the (red) diamonds are
obtained minimizing Eq.~\ref{Eq:en}.  The solid (blue) line shows the
same quantity calculated using the self-consistent approximation
(SCA).  The EFM approach and the SCA give a similar behavior for
$\langle V(r)V(0) \rangle$, characterized by an algebraic,
$\propto 1/r^3$, decay at large distances.  The green solid line shows
a Gaussian approximation which captures much of the quantitative
details of the screened disorder potential correlation function, but
not the power law $1/r^3$ decay.

\begin{figure}
\bigskip
\hspace{0.1\hsize}
\epsfxsize=0.9\hsize
\epsffile{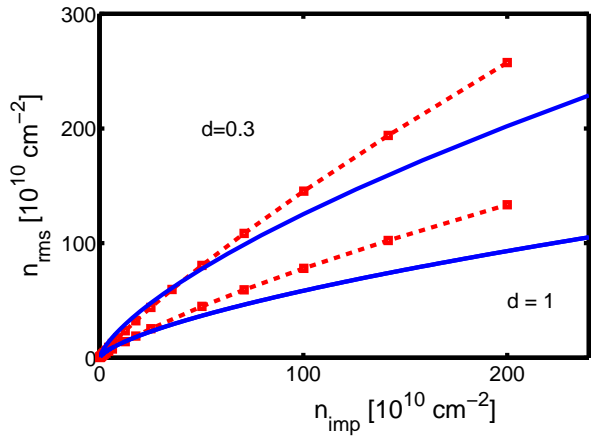}
 \caption{\label{fig:nrms_nimp}  
 	   Results for $n_{\rm rms}$ as a function of $n_{\rm imp}$
	   at the Dirac point for $r_s=0.8$. The red dashed lines are 
	   the results obtained
	   minimizing graphene's energy functional and the solid lines
	   are the SCA results.}
\end{figure}

\begin{figure}
\bigskip
\hspace{0.1\hsize}
\epsfxsize=0.9\hsize
\epsffile{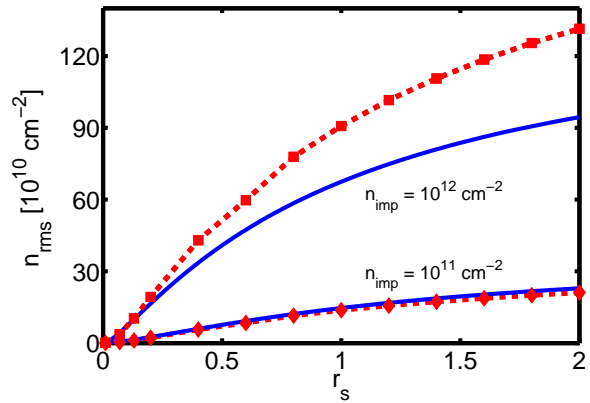}
 \caption{\label{fig:nrms_rs}
 	   Results for $n_{\rm rms}$  as a function of $r_s$
	   at the Dirac point for two values of $n_{\rm imp}$
	   and $d=1$~nm. The red dashed lines are the results obtained
	   minimizing graphene's energy functional and the solid lines
	   are the SCA results.}
\end{figure}

\begin{figure}
\bigskip
\hspace{0.1\hsize}
\epsfxsize=0.9\hsize
\epsffile{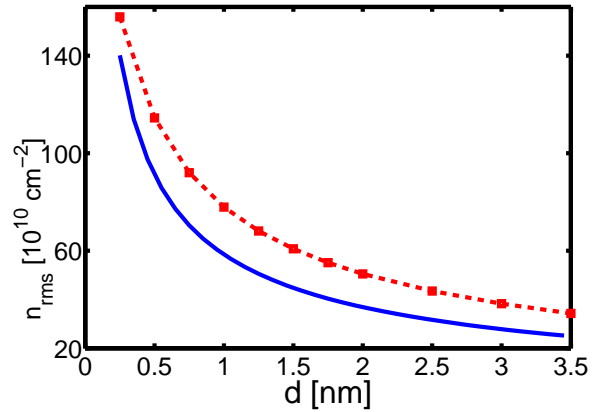}
 \caption{\label{fig:nrms_d}  
            Results for $n_{\rm rms}$  as a function of $d$
	   at the Dirac point for $n_{\rm imp}=10^{12}$~cm$^{-2}$
	   and $r_s=0.8$. The red dashed lines are the results obtained
	   minimizing graphene's energy functional and the solid lines
	   are the SCA results.}
\end{figure}

In Figs.~\ref{fig:nrms_nimp}\hspace{-0.05in}
  % ,fig:nrms_rs,
-\ref{fig:nrms_d}, $n_{rms}$ at the Dirac point is shown as function
of $n_{imp}$, $r_s$ and $d$ respectively.  The red dashed lines show
the results obtained using the EFM theory including exchange, the
solid blue lines are the results obtained using the SCA theory.  In
general the SCA gives values of $n_{rms}$ smaller than the EFM theory
but in general there is good semi-quantitative agreement especially at
low $r_s$ and $n_{imp}$.

\section{Graphene Conductivity}
\subsection{High-density: Boltzmann Transport Theory}
\label{Sec:Boltzmann}

In this section we investigate the graphene transport
for large carrier densities ($n \gg n_i$), where the system is
homogeneous.
We show in detail the microscopic transport
properties at high carrier density using the Boltzmann transport
theory.~\cite{kn:dassarma1999}  We calculate the
conductivity $\sigma$ (or mobility $\mu = \sigma/ne$) in the
presence of randomly distributed Coulomb impurity 
charges near the surface with the electron-impurity interaction being
screened by the 2D electron gas in the random phase approximation
(RPA). Even though the screened Coulomb scattering is the most important
scattering mechanism in our calculation, there are additional
scattering mechanisms (i.e. neutral point defects) unrelated to the
charged impurity  scattering  for very high mobility samples.
Point defects gives rise to a constant conductivity  in contrast to 
charged impurity scattering which produces a conductivity linear
in $n/n_i$.  Our formalism can include both effects, where zero
range scatterers are treated with an effective point defect density of 
$n_p$.  For the purpose of this calculation, we neglect all phonon 
scattering effects, which were considered recently with the finding 
that acoustic phonon scattering gives rise to a resistivity that is 
linear in temperature.~\cite{kn:hwang2008}

We start by assuming graphene to be a homogeneous 2D carrier system
of electrons (or holes) with a carrier density $n$ induced by the
external gate voltage. The low-energy band Hamiltonian for homogeneous graphene is 
well-approximated by a 2D Dirac equation for massless particles, 

\begin{equation}
H = \hbar v_F (\sigma_x k_x + \sigma_y k_y),
\end{equation}
where $v_F$ is the 2D Fermi velocity, $\sigma_x$ and $\sigma_y$ are Pauli spinors 
and ${\bf k}$ is the momentum relative to the Dirac points.
The corresponding eigenstates are given by the plane wave 
 $\psi_{s {\bf k}}({\bf r}) = \frac{1}{\sqrt{A}}\exp(i {\bf k} \cdot {\bf r}) 
F_{s {\bf k}}$,  
where $A$ is the area of the system, $s = \pm 1$ indicate the
conduction ($+1$) and valence ($-1$) bands, respectively, and
$F_{s {\bf k}}^{\dagger} = \frac{1}{\sqrt{2}}(e^{i\theta_{\bf k}},s)$
with $\theta_{ {\bf k}} = \tan(k_y/k_x)$ being the polar angle of the
momentum $\hbar {\bf k}$.
The corresponding energy of graphene for 2D wave vector ${\bf k}$ is given by 
$\epsilon_{s {\bf k}}= s \hbar v_F | {\bf k}|$, and 
the density of states (DOS) is given
by $D(\epsilon) = g|\epsilon|/(2\pi \hbar^2 v_F^2)$, where $g=g_s g_v$
is the total degeneracy ($g_s = 2, g_v = 2$ being the spin and valley
degeneracies, respectively). 

When the external force is  weak and the displacement of the
distribution function from the thermal equilibrium value is small, we
can use linearized Boltzmann equation within relaxation time
approximation.  In this case the conductivity for graphene can be written by

\begin{equation}
\sigma = \frac{e^2 v_F^2}{2} \int d\epsilon_k D(\epsilon_k)
\tau(\epsilon_k) \left ( 
  -\frac{\partial f(\epsilon_k)}{\partial \epsilon_k} \right ).
 \label{eq:sigma}
\end{equation}

Note that $f(\epsilon_k)$ is the Fermi distribution function, 
 $f(\epsilon_k) =\{ 1+\exp[(\epsilon_k-\lambda)]/k_B T \}^{-1}$ 
where the finite temperature chemical potential $\lambda(T)$ is
determined  self-consistently to conserve the total number of 
electrons.  At $T=0$,
$f(\epsilon)$ is a step function at the Fermi energy $E_F \equiv
\lambda(T=0)$, and we then recover the Einstein relation 
$\sigma = \frac{e^2 v_F^2}{2} D(E_F)\tau(E_F)$.
In Eq. (\ref{eq:sigma}) $\tau(\epsilon_{s {\bf k}})$ is the relaxation time or
the transport scattering time of the collision and is given by

\begin{eqnarray}
\frac{1}{\tau(\epsilon_{s{\bf k}})} & = &\frac{2\pi}{\hbar} \sum_a
n_i^{(a)} \int \frac{d^2 k'}{(2\pi)^2} | \langle V_{s{\bf k},s{\bf k}'}^{(a)}\rangle |^2  
\nonumber \\ 
 &\times & [1-\cos\theta_{{\bf k}{\bf k}'}]
\delta\left (\epsilon_{s{\bf k}} - \epsilon_{s{\bf k}'} \right ),
\label{eq:tk}
\end{eqnarray}
where $\theta_{{\bf kk}'}$ is the scattering angle between the scattering in-
and out- wave vectors ${\bf k}$ and ${\bf k}'$, and  
$\langle V_{s{\bf k},s'{\bf k}'}^{(a)} \rangle$ is the matrix element of the
scattering potential associated with impurity disorder in the
graphene environment, and $n_i^{(a)}$ is the number of impurities per unit
area of the $a$-th kind of impurity.  Note that since we consider the elastic impurity scattering
the interband processes ($s \neq s'$) are not permitted.   

The matrix element of the scattering potential
of randomly distributed screened impurity charge centers in graphene
is given by 
\begin{equation}
 |\langle V_{s{\bf k},s{\bf k}'}^{(a)} \rangle |^2 = \left |
   \frac{V_i(q,d)}{\varepsilon(q)} \right |^2 \frac{1+\cos \theta}{2}
 \label{eq:vkk}
 \end{equation} 
 where $q = |{\bf k} - {\bf k}'|$, $\theta \equiv \theta_{{\bf k} {\bf k}'}$, and
 $V_i(q,d)=2\pi e^2\exp(-qd)/(\kappa q)$ is the Fourier transform 
 of the 2D Coulomb potential in an effective background lattice
 dielectric constant $\kappa$, where $d$ is the location of the charged
 impurity measured from the graphene sheet. The factor $(1+\cos\theta)/2$
 arises from the sublattice symmetry (overlap of wave function).~\cite{kn:ando2006}

In Eq.~(\ref{eq:vkk}), $\varepsilon(q) \equiv \varepsilon(q,T)$ is the 2D finite 
temperature static RPA dielectric (screening) function appropriate for 
graphene, given by  $\varepsilon(q,T) = 1 + v_c(q) [1-G(q)] \Pi(q,T)$, where $\Pi(q,T)$ is 
the graphene irreducible finite-temperature polarizability 
function,~\cite{kn:hwang2008} $v_c(q)$ is the Coulomb interaction, 
and $G(q)$ is the local field correction.  In RPA, $G(q)=0$ and in 
Hubbard approximation (HA), 
$G(q) = 1/(g_s g_v) \times  (q / \sqrt{q^2 + k_F^2})$.~\cite{kn:jonson1976} 
% On the other hand, the matrix element of the scattering potential for neutral zero range scatterers 
% is given by 
%  \begin{equation}
%  |\langle V_{s{\bf k},s{\bf k}'}^{(a)} \rangle |^2 =  
%   \left | V_0 \right |^2 \frac{(1+\cos \theta)}{2}.
%  \end{equation} 

\begin{figure}
\bigskip
\hspace{0.1\hsize}
\epsfxsize=0.9\hsize
\epsffile{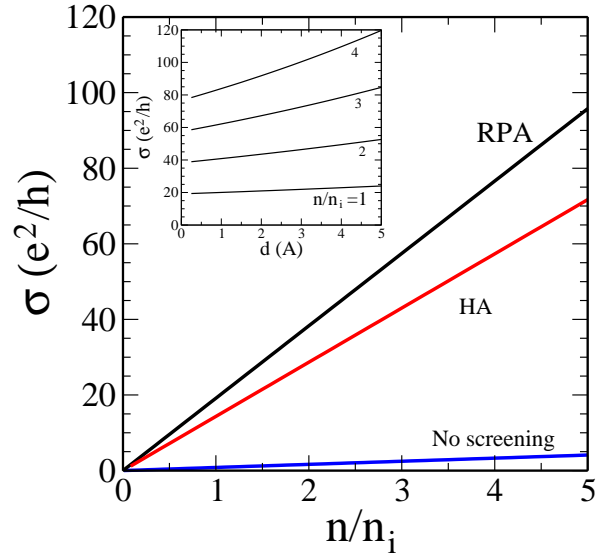}
\caption{\label{Fig:ssc1} 
Calculated graphene conductivity as a function of 
carrier density ($n/n_i$, where $n_i$ is the impurity density)
limited by Coulomb scattering.
RPA (HA) indicates the result with the RPA (HA) screened Coulomb
scattering, and ``no screening'' indicates the results with bare Coulomb
scattering.  Note that the calculated conductivity with
unscreened Coulomb potential is less
than $4e^2/h$ for given density range.
In inset the effect of remote scatters is shown.
Here $d$ is the distance between the 2D graphene layer and
the 2D impurity layer.}
\end{figure}

In Fig.~\ref{Fig:ssc1} we show the
calculated graphene conductivities limited by screened charged
impurities. The RPA screening used in our calculation is
the main approximation. We also show results for HA
screening~\cite{kn:jonson1976} which includes local field corrections
approximately. We note that since the graphene is a weakly
interacting system ($r_s <1$) the correlation effects are not strong.
We emphasize that in order to get quantitative agreement with
experiment, the screening effects must be included. Using the  
unscreened dielectric function would have conductivity less
than $4e^2/h$ for the entire range of gate voltages used in the
experiment.
Our main result with screened Coulomb impurities is the quantitative
agreement with 
experiments in the regime where the conductivity is linear in density.
In inset we show the effect of remote scatterers which are
located at a distance $d$ from the interface.  The main effect of
remote impurity scatterings is that the conductivity deviates from
the linear behavior with density and
increases with both the distance $d$ and $n/n_i$.
 
\begin{figure}
\bigskip
\hspace{0.1\hsize}
\epsfxsize=0.9\hsize
\epsffile{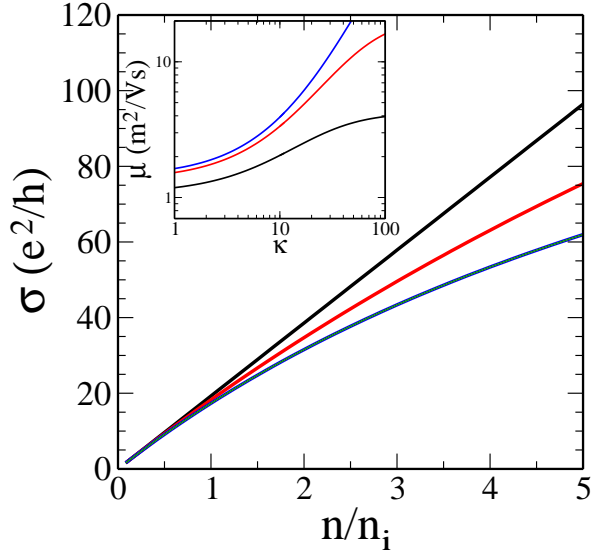}
\caption{\label{fig:ssc2} Graphene conductivity calculated using
a combination of short and long range scatterers. 
In this calculation, we use $n_p/n_i=0$, 0.01, 0.02 (top to bottom).
In inset we show the graphene mobility as a function of dielectric
constant ($\kappa$) of substrate for different carrier densities
$n = 0.1$, 1, $5\times 10^{12}$cm$^{-2}$ (from top to bottom) 
in the presence of both long ranged
charged impurity ($n_i = 2\times 10^{11}$ cm$^{-2}$) and
short-ranged neutral impurity ($n_p = 0.4\times 10^{10}$ cm$^{-2}$). 
$V_0=1$ KeV\AA$^2$ is used in this calculation, which corresponds
to the Coulomb potential of electron density $n= 10^{12}$ cm$^{-2}$.}
\end{figure}

\begin{figure}[t!]
\bigskip
\hspace{0.1\hsize}
\epsfxsize=0.9\hsize
\epsffile{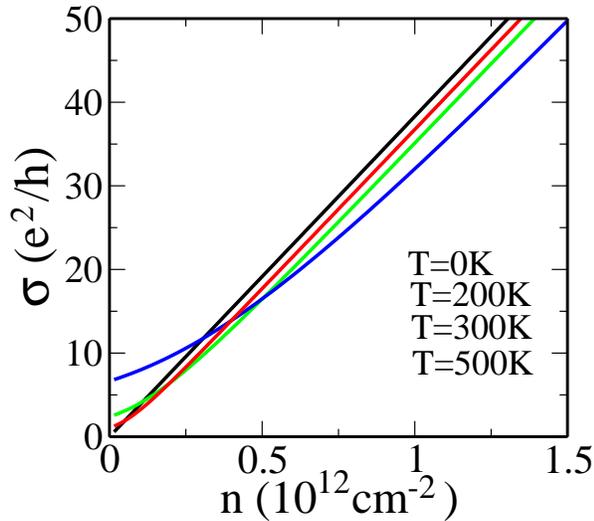}
\caption{\label{fig:ssc3} 
Calculated conductivity for different temperatures $T=0$, 200, 300,
500K (top to bottom) as a function of density with
an impurity density $n_i=5\times 10^{11}$cm$^{-2}$.}
\end{figure}

 For very high mobility samples, 
one finds a sub-linear conductivity instead of
the linear behavior with density.  Such 
high quality samples presumably
have a small charge impurity concentration $n_i$ and 
it is therefore likely that  point defects here play a more dominant role.  
Point defects 
gives rise to a constant conductivity in contrast to charged
impurity scattering which produces a conductivity linear
in $n/n_i$.  
In the presence of both the long-ranged charged
impurity and the short-ranged  neutral impurity, 
the total scattering time becomes
${1}/{\tau_t} = {1}/{\tau_{i}} + 1/\tau_{0}$ 
where $\tau_i$ ($\tau_{0}$) is the scattering time due to 
charged Coulomb (short ranged) impurities. 
Shown in
Fig.~\ref{fig:ssc2} is the graphene conductivity calculated
including both charge impurity  and zero range point defect
scattering for different ratios of the point scatterer
impurity density $n_p$ and the charge impurity density $n_i$.  For
small $n_p/n_i$  we find the linear conductivity that
is seen in most experiments and for large $n_p/n_i$  we see the
flattening out of the 
conductivity curve (which in the literature~\cite{kn:zhang2005} has been
referred to as the sub-linear conductivity). 
We believe this
high-density flattening of the graphene conductivity
is a non-universal crossover behavior arising from the competition
between two kinds of scatterers. In general this crossover occurs when
two scattering potentials are equivalent, that is, $n_iV_i^2  \approx
n_p V_0^2$.  

In the inset of Fig.~\ref{fig:ssc2} we show our calculated
mobility in the presence of both charged 
impurities and short-ranged impurities. 
As the scattering limited by the short-ranged impurity dominates over that
by the long-ranged impurity (e.g. $n_p V_0^2 \gg n_iV_i^2$)   
the mobility is no longer
dependent on the charged impurity and approaches its limiting value
\begin{equation}
\mu = \frac{e}{4\hbar}\frac{(\hbar v_F)^2}{n}\frac{1}{n_{p}V_0^2}.
\end{equation}
The limiting mobility depends only on neutral impurity concentration 
$n_{p}$ and carrier density, which indicates that to get high
graphene mobility it 
is necessary to have defect free graphene.

 Finally in Fig. \ref{fig:ssc3} we show the calculated temperature
dependent conductivity for different temperatures as a function of
density. 
We note that there are two independent sources of temperature
dependent resistivity in our calculation. One comes from the energy
averaging defined in Eq. (\ref{eq:sigma}), 
and the other is the explicit temperature dependence of the 
dielectric function $\varepsilon(q,T)$  which
produces a 
temperature dependent $\tau(\varepsilon,T)$.  Figure~\ref{fig:ssc3} shows 
that in the high density 
limit the conductivity decreases as the temperature increases, but
in the low density limit the conductivity shows non-monotonic behavior,
i.e. $\sigma(T)$ has a local minimum at a finite temperature and
increases as the temperature increases.
Thus, we  find that the calculated conductivity shows a
non-monotonicity in the low density limit, i.e., at low temperatures 
the conductivity shows metallic behavior and at high temperatures
it shows insulating behavior. The
non-monotonicity of temperature dependent $\sigma(T)$ is understood
to arise from temperature dependent screening.~\cite{kn:hwang2008b}
We mention that for $T\gtrsim 100$~K, phonons contribute to the
temperature dependence of graphene 
conductivity.~\cite{kn:hwang2008,kn:chen2008b}

\subsection{Low Density: Effective Medium Theory}
\label{Sec:EMT}
\begin{figure}[t!]
\bigskip
\hspace{0.1\hsize}
\epsfxsize=0.9\hsize
\epsffile{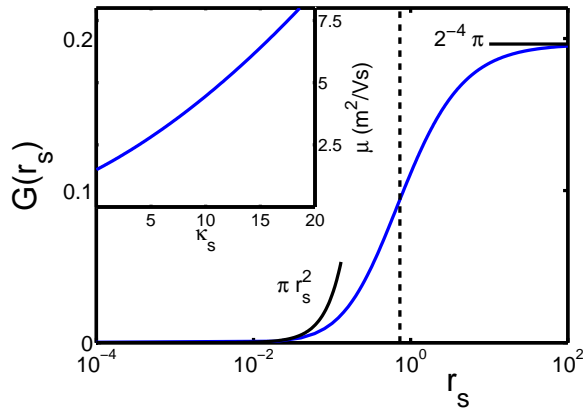}
\caption{\label{Fig:cond_rs} 
The main panel shows $G[r_s]$ that parameterizes the inverse scattering time
in the Boltzmann theory.  The analytic form of $G[r_s]$ can be found in 
Ref.~\onlinecite{kn:adam2007a}.  The dashed line shows the experimentally 
relevant regime for graphene on SiO$_2$ substrates.  Inset shows
sample mobility as a function of substrate dielectric constant $\kappa_s$ 
for $n_{\rm imp} = 2 \times 10^{11}~cm^{-2}$.  Changing $\kappa_s$ by 
a factor of 2, increases mobility by 50 percent. }
\end{figure}

At low density the fluctuations in carrier density become larger than
the average density.  To understand the transport properties of this
inhomogeneous system, Ref.~\onlinecite{kn:rossi2008b} developed an
effective medium theory where graphene's conductivity is found by
solving an integral equation
\begin{eqnarray}
\int dn \frac{\sigma[n] - \sigma_{\rm EMT}}{\sigma[n] + \sigma_{\rm EMT}} P[n] = 0 
\label{Eq:EMT}
\end{eqnarray}
where $P[n]$ is the density distribution function and $\sigma[n]$ is
the (local) Boltzmann conductivity discussed in
Sec.~\ref{Sec:Boltzmann}.  For the purpose of this section, we take
$\sigma_B[n] = (2e^2/G[r_s] h) n/n_{\rm imp}$ where $G[r_s]$ was derived
in Ref.~\onlinecite{kn:adam2007a} and shown in Fig.~\ref{Fig:cond_rs}.
The quantitatively accurate theory using $P[n]$ derived from the EFM
of Sec~\ref{SubSec:EFM} was developed in
Ref.~\onlinecite{kn:rossi2008b}.  Here we derive analytical results
obtained by using model distribution functions for $P[n]$, which as
discussed in Ref.~\onlinecite{kn:rossi2008b} show quantitative
agreement with the numerical theory only for small $r_s$ and low
$n_{\rm imp}$.  To illustrate this method, we first consider $P[n]$ to
be a Gaussian distribution.  Requiring that $\int n^2 P[n] = n_{\rm
rms}^2$ fixes all the free parameters.  Solving Eq.~\ref{Eq:EMT} then
gives $ z \exp^{-z^2}(\pi {\rm Erfi}[z] -{\rm Ei}[z^2]) = \sqrt{\pi}/2
$, where ${\rm Erfi}$ is the imaginary error function, ${\rm Ei}$
is the exponential integral function and $z = \sigma_{\rm
EMT}/(\sqrt{2} \sigma_B[n_{\rm rms}]) \approx 0.405$ giving $\sigma_{\rm
EMT} \approx 0.9925~ \sigma_{\rm SCA}$, where we use the results of
Sec.~\ref{SubSec:SCA} that $\sigma_{\rm SCA} = \sigma_B[n^*]$ 
and $n_{\rm rms} \approx \sqrt{3} n^*$.  The
development of an effective medium theory for
graphene~\cite{kn:rossi2008b} now allows us to reinterpret the
results of Ref.~\onlinecite{kn:adam2007a} as equivalent to the assumption that
$P[n]$ is Gaussian with density fluctuations determined by the
self-consistency condition $E_{\rm F}^2 = \langle V_{D} \rangle^2$.                

We can explore other functional forms for $P[n]$.  For a Lorentzian
$P[n] = (n_L/\pi)/(n_L^2 + n^2)$, one can solve Eq.~\ref{Eq:EMT}
analytically giving $\sigma_{\rm EMT} = \sigma_B[n_L]$.  If one
identifies the width of the Lorentzian with the self-consistent
carrier density $n_L = n^*$, then this provides another way to
understand the self-consistent transport result.  Subsequent to the
results of Ref.~\onlinecite{kn:rossi2008b}, Fogler developed an
effective medium theory using $P[n] = (1/\sqrt{2}~n_{\rm rms}) \exp[
-\sqrt{2} |n|/n_{\rm rms}]$, where similar to the Gaussian
distribution case, requiring normalization and setting 
$\langle n^2 \rangle = n_{\rm rms}^2$ fixes all the free parameters.  Solving
Eq.~\ref{Eq:EMT} gives $z e^z \Gamma[z] = 1/2$ where $\Gamma$ is the
Gamma function and $z = \sqrt{2} \sigma_{\rm EMT}/\sigma_B[n_{\rm rms}]
\approx 0.610$.  Again, approximating $n_{\rm rms} \approx \sqrt{3}
n^*$, we find $\sigma_{\rm EMT} \approx 0.75~\sigma_{\rm SCA}$, which
is different from the numerical results obtained in
Ref.~\onlinecite{kn:fogler2008b}.  While these analytical approximations
are useful in providing a qualitative understanding of graphene
transport, quantitative differences remain between these and the full
numerical solution~\cite{kn:rossi2008b} especially at large impurity
concentrations and large $r_s$ (See e.g. Fig.~\ref{fig:nrms_rs}).

\subsection{Suspended Graphene}
\label{Sec:Suspended}
One of the direct consequences of charged impurity scattering
in graphene is the prediction~\cite{kn:hwang2006c} that 
the elimination of charged impurities from the graphene
environment, for example, by suspending graphene, without
any substrate would lead to a much enhanced carrier mobility.
Recently, Bolotin et al.~\cite{kn:bolotin2008,kn:bolotin2008b}
managed to remove charged impurities from the graphene
environment by suspending graphene without any substrate (and simply
current-annealing away any remnant impurities on the 
graphene surface).  This immediately led to an 
oder of magnitude increase (to $\mu \sim 10^{5}~{\rm cm}^2/{\rm Vs}$)
in the graphene mobility as predicted theoretically.  Recent
theoretical work~\cite{kn:adam2008b} shows excellent
agreement with the transport measurements on 
suspended graphene,~\cite{kn:bolotin2008,kn:bolotin2008b,kn:du2008}
with both the reduced impurity density and the modified
screening (due to the elimination of the substrate) contributing
to the graphene conductivity.  Since phonon 
scattering effects in graphene are weak upto room 
temperature,~\cite{kn:hwang2008} the enhanced graphene mobility
arising from the elimination of charged impurities may lead
to very high ($> 10^5~{\rm cm}^2/{\rm Vs}$) graphene mobilities
even at room temperature. 

\section{Discussion of experiments}
\label{Sec:Expts}

\begin{figure}
\bigskip
\hspace{0.1\hsize}
\epsfxsize=0.8\hsize
\begin{center}
\epsffile{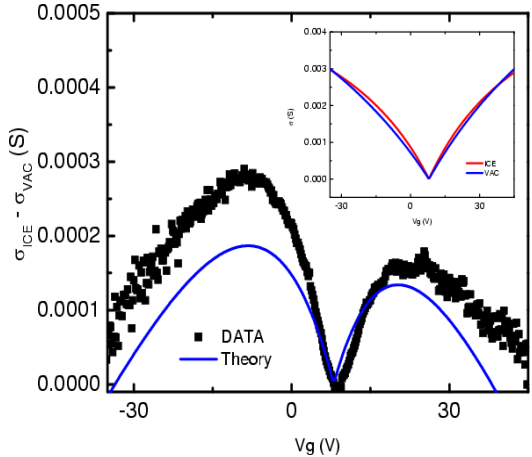}
\end{center}
\caption{\label{Fig:ice_expts} Data taken from 
Ref.~\protect{\onlinecite{kn:jang2008}} shows the effect of dielectric
screening on graphene.  Data points show the difference in graphene
conductivity before and after depositing ice, while solid line shows
the theory with no adjustable parameters.  The non-monotonic behavior
is a consequence of the competing effects of dielectric screening on
Coulomb and short-range scatterers, whereas the quantitative agreement
is a stringent test of the the Boltzmann theory.  Inset, also taken
from Ref.~\protect{\onlinecite{kn:jang2008}} shows the raw
experimental data.}
\end{figure}

One of the very first puzzles in graphene transport
experiments~\cite{kn:novoselov2004} was that the conductivity
was linear in carrier density, whereas existing theory~\cite{kn:shon1998}
predicted constant conductivity at high density.  As discussed in the 
introduction, the linear in density emerges naturally from the
Boltzmann transport theory of charged 
impurities,~\cite{kn:ando2006,kn:nomura2006a,kn:cheianov2006,kn:hwang2006c,kn:adam2007a} and to our knowledge, no other theory produces this linear
behavior without a fine-tuning of parameters to make the scattering
mimic Coulomb impurities.  Moreover, three recent experiments have
rigorously verified the high density predictions of the Boltzmann
transport theory.  In Ref.~\onlinecite{kn:tan2007} sample mobility was
correlated with shift of the Dirac point and plateau width showing
qualitative and semi-quantitative agreement with the theory presented
here.  Reference~\onlinecite{kn:chen2008} were able to directly
measure the effect of Coulomb scatterers by intentionally adding
potassium ions to graphene in ultra-high vacuum observing
qualitatively all the predictions of the (self-consistent) Boltzmann
theory.  Finally, Ref.~\onlinecite{kn:jang2008} was able to tune graphene's
fine structure constant by depositing ice on top of graphene, and thereby 
precisely testing predictions of the Boltzmann theory.  Results from
this experiment are shown in Fig~\ref{Fig:ice_expts}.  Since the
dielectric constant of the SiO$_2$ substrate and that of ice are known, there
are no adjustable parameters in the theoretical curve.

\section{The Minimum Conductivity Puzzle}
\label{Sec:MinCond}

As described above in Sec.~\ref{Sec:EMT}, our recent 
theoretical work~\cite{kn:adam2007a,kn:rossi2008b} provides
a satisfactory explanation for the minimum conductivity
phenomenon in graphene near the charge neutrality (i.e. Dirac) 
point.  In particular, early theoretical work~\cite{kn:fradkin1986,kn:ludwig1994,kn:shon1998,kn:katsnelson2006,kn:tworzydlo2006} predicted
a universal $T=0$ minimum conductivity $\sigma_{\rm min} = 4e^2/\pi h$ 
at the graphene Dirac point in clean disorder-free systems.  The inclusion
of disorder-induced quantum anti-localization effect, assuming
no intervalley scattering, leads to a theoretical infinite minimum 
conductivity at the Dirac point, whereas the presence of inter-valley
scattering localizes the system leading to zero conductivity at the 
Dirac point.  This confusing theoretical picture stands in stark 
contrast to the experimental reality, where the graphene 
conductivity is approximately a constant (as a function of gate
voltage or carrier density) around the Dirac point, with 
this constant minimum conductivity plateau having a non-universal
sample dependent value ($\sim 4e^2/h - 20 e^2/h$).

It was first suggested in Ref.~\onlinecite{kn:hwang2006c} that 
the minimum conductivity phenomenon  is closely related to 
the break-up of the graphene landscape into inhomogeneous
puddles of electrons and holes around the Dirac point due to the
effect of the charged impurities in the environment.  This physical
idea was further developed into a quantitatively successful 
theory (see Sec.~\ref{Sec:EMT} above) in 
Refs.~\onlinecite{kn:adam2007a,kn:rossi2008b}, where it was 
shown that a self-consistent treatment of the impurity
induced electron-hole puddles coupled with the Boltzmann 
transport theory provides excellent description of the non-universal
behavior of the minimum conductivity around the Dirac point.  In 
particular, the sample dependence of the minimum conductivity arises
from the different impurity disorder in different samples.

Quantum effective field theories of graphene minimum conductivity, 
which predict a universal minimum conductivity, are inapplicable
to real graphene samples because real graphene is dominated by 
disorder induced inhomogeneity near the Dirac point, which is
outside the scope of the quantum field theories.  Although the
self-consistent effective medium theory developed by us~\cite{kn:adam2007a,kn:rossi2008b} gives reasonable agreement with the experimental observations,
the key conceptual question of what happens at $T=0$ as disorder
also goes to zero still remains open.  Such a scenario 
is of course experimentally irrelevant (since experiments 
are performed at finite temperatures in disordered systems), but the
theoretical question of the conductivity crossover from the 
inhomogeneous Boltzmann regime of Refs.~\onlinecite{kn:adam2007a,kn:rossi2008b}
to the homogeneous quantum transport regime is an interesting open 
question.  Recent attempts to understand such crossover phenomena
include several complementary theoretical avenues (See 
Refs.~\onlinecite{kn:adam2008d,kn:cheianov2007,kn:muller2008} and
references therein).

\section{Recent Work}
\label{Sec:Recent}

Among recent relevant work not discussed in this review we mention
a detailed calculation of the temperature dependent graphene
conductivity due to electron-phonon scattering,~\cite{kn:hwang2008}
a detailed calculation of the temperature dependent 
graphene conductivity due to the temperature 
dependence of the screening of charged impurity
scattering,~\cite{kn:hwang2008b} a calculation of graphene 
density of states as modified by impurity 
scattering,~\cite{kn:hu2008} a consideration of 
percolation induced localization transition in graphene 
nanoribbons,~\cite{kn:adam2008c} a theory of charged 
impurity screening in graphene bilayers,~\cite{kn:hwang2008c}
and a prediction of graphene magnetoresistance induced
by a parallel magnetic field through the spin-polarization
dependence of screening.~\cite{kn:hwang2008d}
 
\section{Conclusion} 
\label{Sec:Conclusion}

We have developed here the theory for Coulomb impurities on graphene.
As we have shown, Coulomb impurities behave qualitatively different
from short-range scatterers such as point defects or missing atoms.
Away from the Dirac point, the physics is well described by a
semi-classical Boltzmann transport theory, while at the Dirac point
density fluctuations dominate breaking the system into puddles of
electrons and holes.  We have shown how these inhomogeneities can be
characterized by a mean-field self-consistent theory and by
numerically minimizing graphene's energy functional, and that an
effective medium theory can be employed to describe the low density
transport properties giving semi-quantitative agreement with 
experimental results.

We thank Michael Fuhrer for valuable discussions.  This work is 
supported by US-ONR and NSF-NRI-SWAN.

\vspace{-0.2in}
% \bibliography{/Users/shaffiqueadam/Documents/LaTexFiles/Mac_shaffiquebib.bib}
% \bibliography{/home/sadam1/LaTex/shaffiquebib.bib}

\end{document}